\documentclass[11pt]{article}
\usepackage[utf8]{inputenc}
\usepackage{hyperref}
\usepackage{setspace}
\usepackage{palatino}
\usepackage{graphicx}
\usepackage{float}
\usepackage{titling}
\usepackage{multirow}
\usepackage{lscape}
\usepackage{amsmath}
\usepackage{amssymb}
\usepackage{subcaption}
\usepackage{tcolorbox}
\usepackage{lipsum} 
\usepackage[a4paper, total={6in, 9.5in}]{geometry}
\fontfamily{ppl}\selectfont 

\usepackage[american]{babel}

\usepackage{setspace}

\usepackage{rotating}
\usepackage{tabularx}
\newcolumntype{C}{>{\centering\arraybackslash}X}
\newcolumntype{R}{>{\raggedright\arraybackslash}X}
\newcolumntype{L}{>{\raggedleft\arraybackslash}X}
\usepackage{authblk}

\usepackage{booktabs}

\newcounter{suppfigure}

\title{Global Patterns of Knowledge: Language, Genre, and the Geography of Knowledge
\thanks{This article presents preliminary findings from ongoing research. It is subject to revision, and comments are welcome.}
}

\author{
Akira Matsui\textsuperscript{1,*}, 
Fujio Toriumi\textsuperscript{2}, 
Mitsuo Yoshida\textsuperscript{3}, 
Taichi Murayama\textsuperscript{4}, 
Shiori Hironaka\textsuperscript{5}\\
\textsuperscript{1}Kobe University \\
\textsuperscript{2}The University of Tokyo \\
\textsuperscript{3}University of Tsukuba \\
\textsuperscript{4}Yokohama National University \\
\textsuperscript{5}Kyoto University \\
\textsuperscript{*}Corresponding author: \href{mailto:amatsui@rieb.kobe-u.ac.jp}{amatsui@rieb.kobe-u.ac.jp}
}

\date{\today}

\begin{document}
{
\setstretch{.8}
\maketitle
\begin{abstract}
Online platforms, particularly Wikipedia, have become critical infrastructures for providing diverse linguistic and cultural contexts. This human-curated knowledge now forms the foundation for modern AI. However, we have not yet fully explored how knowledge production capability vary across languages and domains. Here, we address this gap by applying economic complexity analysis to understand the editing history of Wikipedia platforms. This approach allows us to infer the latent mode of ``knowledge-production'' of each language community from the diversity and specialization of its contributed content. We reveal that different language communities exhibit distinct specializations, particularly in cultural subjects. Furthermore, we map the global landscape of these production modes, finding that the structure of knowledge production strongly reflects geopolitical boundaries. Our findings suggest that while a common mode of knowledge production exists for standardized topics such as science, it is more diverse for cultural topics or controversial subjects such as conspiracy theories. The association between differences in knowledge production capability and geopolitical factors implies how linguistic and cultural dynamics shape our worldview and the biases embedded in Wikipedia data, a unique, massive, and essential dataset for modern AI.
\end{abstract}
}

\section{Introduction}
Humans, as information-seeking creatures, expand their ability to search for and accumulate information through online platforms~\cite{DaleArchitectural2024}. Online platforms also provide collaborative environments where individuals cooperate in acquiring and organizing information to produce knowledge~\cite{Badashian2014Involvement, Vasilescu2013StackOverflow, Dabbish2012, Kim2015Understanding}. 
The knowledge production structure of a given language depends on its linguistic, cultural, or geopolitical contexts~\cite{kramsch2014language,sharifian2017cultural,alkhamissi2024investigating, li2024land}. 
Because knowledge is documented in language, cross-lingual interactions play a pivotal role~\cite{Hale2013Multilinguals, Kim2015Understanding}. Cross-cultural and linguistic borders can shape the co-evolution and similarity of knowledge across different languages~\cite{Samoilenko2016Linguistic}.
Disentangling these interconnections can enhance not only our understanding of the world but also today's AI models.
The data retrieved from online platforms or the World Wide Web, in general, is essential for AI training, including Large Language Models (LLMs), and online platforms constitute the de facto infrastructure of modern society for knowledge production and accumulation.

The knowledge production structure within a linguistic-cultural sphere mirrors the cultural priorities and information disparities embedded globally~\cite{Nordlund1998The, Altarriba2022The}, a phenomenon particularly prominent in English~\cite{Piller2022Peripheral}. 
This underscores the need for empirical investigations into how such modes can embed inherent biases into data, but analyzing these global patterns is not a trivial task.
First, similar to international trade, languages may hold a comparative advantage over others in knowledge production.
For example, although many documents about Sushi are available on the internet in various languages, Japanese-language versions have a significant comparative advantage because the dish originated in Japan.
While it has a comparative advantage over others, this does not mean other languages do not maintain or produce such knowledge, as they often require their own language versions of the original content.
A binary classification of whether a language has particular knowledge does not capture this relationship~\cite{Samoilenko2016Linguistic}; instead, we must study the amount of labor required to maintain that body of knowledge.
Second, knowledge production requires collaboration~\cite{Beers2005Computer}, and this cooperation occurs not only among users within each language but also across languages~\cite{Kim2015Understanding,ElKomboz2024Virtually,Hale2013Multilinguals,Park2021MultilingualWikipedia}.
To understand the characteristics of this complex system, we must look beyond individual editing activities to large-scale, structural patterns.

Moreover, language is not the only factor that creates structural differences in knowledge production capability across languages.
The domains of knowledge that producers create are one of the primary principles~\cite{Driscoll2019Genre,Olinghouse2012The}, as access patterns vary among domains (topics) and countries~\cite{piccardi2023curious}.
A domain like scientific knowledge, for instance, can disseminate globally and rapidly in a standardized manner~\cite{Olinghouse2012The}.
The community can openly validate its validation and description against global standards, as exemplified by scientific papers. 
In contrast, history or controversial topics often link to specific languages and cultural or geopolitical boundaries~\cite{li2024land}, and such knowledge often lacks a standardized mode of description.
Between these typical domains, some knowledge exists along a spectrum. 
While conspiracy theories sometimes use pseudo-scientific or fact-based arguments, they often remain specific to particular cultural spheres, but editors sometimes document them as if they were scientific knowledge. 
A classic example is the ``flat-earth'' theory, a belief once prevalent within a specific historical and cultural framework, which is now not supported by scientific consensus, yet is still discussed as if it were supported by scientific evidence~\cite{uscinski2014american}. 
Knowledge with strong cultural backgrounds, on the other hand, can have standardized production modes.
The documentation of culinary practices, for instance, often aims for transmission across multiple languages and cultural regions~\cite{winata2024worldcuisines,cao2024cultural}. 
This suggests that even a single language can contain different modes in its knowledge production across domains.

To capture such a complex structure, one needs to model the latent knowledge production capabilities across different languages and domains. 
In this paper, we employ Economic Complexity Analysis. 
This framework, developed in economic analysis, reveals the sophistication of a given country's industrial structure. Its effectiveness has been supported by empirical studies~\cite{hidalgo2009building, hidalgo2021economic}.
The fundamental principle of the method is to estimate the complexity of the unobservable capabilities underlying a system from the diversity of its observed outputs and their low ubiquity. 
This principle is not limited to economic systems; it is applicable to other complex systems where the portfolio of outputs reflects underlying capabilities, such as scientific knowledge production~\cite{li2023quantification}, or technological activities~\cite{juhasz2024software}.

Here, we investigate the knowledge production capabilities across diverse domains over decades, utilizing the editing histories of over 150 Wikipedia language editions spanning 20 years.
Wikipedia is one of the largest collective intelligence projects in human history, where editors from diverse linguistic and cultural backgrounds collaborate to produce knowledge~\cite{akira_matsui_throw_2024}.
The platform's creation of multiple language editions for a single topic provides valuable data connecting different linguistic communities.
The multilingual aspects of the Wikipedia platform allow us to study the knowledge production capabilities of the same articles or domains across different languages~\cite{Kim2015Understanding,ElKomboz2024Virtually,Hale2013Multilinguals,Park2021MultilingualWikipedia}.
Understanding the linguistic and cultural relevance embedded within the Wikipedia platform is essential to tackling this challenge.
The genres that receive emphasis reflect the cultural and regional realities faced by each language's editor community.

Through this approach, we answer two core questions. First, we quantitatively demonstrate the interplay among languages and domains in their knowledge production capabilities, showing that the capabilities for some domains are uniform across languages, while others are not.
Second, we describe the map of production modes across languages and geopolitical boundaries to examine if the production structure reflects geopolitical boundaries.
The empirical investigations presented in this research make several key contributions. We are the first to map the dynamic, genre-dependent landscape of the knowledge economy hidden behind Wikipedia. This reveals the division of labor in global knowledge production and the cultural and geopolitical factors that shape its structure.
Our analysis provides implications for understanding the structure of knowledge that shapes our digital world and the biases that one of the most popular AI training datasets embeds.

\section{Results}

\subsection{Editorial Histories and Editor Behavior}

We collect Wikipedia editing history data from 2001 to 2024 for over 150 language editions, excluding bot contributions. Building on the publicly available MediaWiki API, we identified articles in each language edition that appear in at least one of the high-level genres we set for the analysis: \texttt{Conspiracy}, \texttt{Wikipedia Controversy}, \texttt{Cooking}, and \texttt{Science} (Method Sec.~\ref{sec:method_topics}). The full set of editing histories across the four genres encompasses over 36.5 million users and around 450,000 titles. 

The dataset contains several statistical regularities that span multiple language editions. 
The editing labor expended by editors is imbalanced, as demonstrated by the Lorenz curves of the total number of edits (Figure~\ref{fig:lorenz_curve}), showing that a small group of editors dominates each genre. This finding aligns with prior research~\cite{Iba2010Analyzing, Lerner2017The, Iiguez2014Modeling}, and we show that this imbalance is not specific to these genres. On the other hand, individual-level editor behavior demonstrates differences among the genres (Figure~\ref{fig:dancer_score_rate}) and the probability of edits being reverted correlates with editor engagement, measured by their number of edits (Figure~\ref{fig:revert_rate_comp_bar}), especially in the Conspiracy Theory and Controversy genres.
In contrast, the other two genres exhibit the opposite trend, with high-engagement editors facing fewer reversions, reflecting distinct editing norms and disputes.

\subsection{Knowledge Production Capabilities Across Languages and Domains}
The aforementioned findings indicate that although the four genres share certain statistical patterns in editing behavior, editors in each genre also exhibit distinctive preferences and editing practices.
Such language-specific differences may suggest that they produce knowledge in different ways depending not only on language but also on domains.
To understand how these differences translate into variations in knowledge production across language editions, we utilize economic complexity analysis~\cite{hidalgo2021economic, hidalgo2009building}, originally developed for international trade. Conceptually, languages act like ``countries'' and article sets like ``products,'' allowing us to quantify how each language’s editorial focus overlaps with others.
By adopting this method for the dataset, we assess the regularities by which Wikipedia language editions specialize in particular genres and overcome the challenges we discussed in the introduction.

\begin{figure}[htbp]
    \centering
    \includegraphics[width=\linewidth]{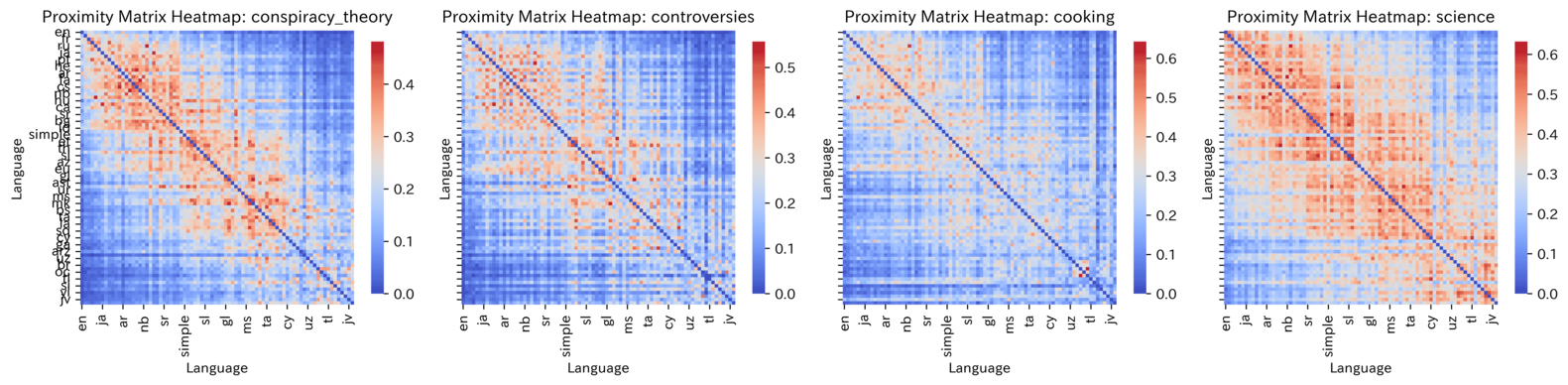}
    \caption{Heatmap of Pairwise Editorial Portfolio Similarity of each genre. The color represents the cosine similarity of editorial specialization between two languages; warmer colors indicate a higher degree of co-specialization on the same set of articles.}
    \label{fig:sim_portfolio_genre_heatmap}
\end{figure}

The analysis reveals coalitions among Wikipedia language editions, where certain editions focus on editing a similar set of articles. First, we calculate similarities in the editing portfolios of 80 language editions for each genre (Figure~\ref{fig:sim_portfolio_genre_heatmap}, Method Sec.~\ref{sec:method_complexity}). This simple analysis reveals two groups of Wikipedia editions. These groups exhibit high internal similarity but low similarity to each other.
The first group in the upper left comprises major European and Asian language editions, whereas the lower right group contains editions of smaller European, South Asian, or Middle Eastern languages (e.g., Croatian, Estonian).
This distinction is particularly evident in the Conspiracy Theory genre, as shown on the left of Figure~\ref{fig:sim_portfolio_genre_heatmap}. In contrast, the Science genre exhibits high similarity across most language pairs, indicating that its articles are edited more uniformly than those in other genres.

\begin{figure}[htbp]
    \centering
    \includegraphics[width=1\linewidth]{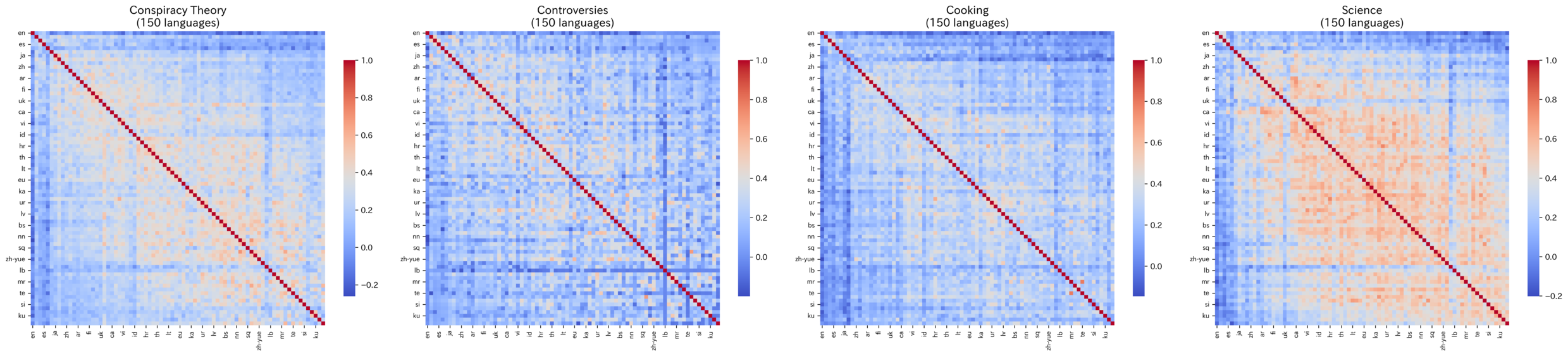}
    \caption{Heatmap of the Language RCA Similarity for Each Genre. Each heatmap visualizes the language's RCA similarity of knowledge production, calculated by the correlation between the RCA values of each language pair. Warmer colors represent higher values, suggesting that two languages have similar specializations in their knowledge production.}
    \label{fig:heatmap_rca_combined_conspiracy}
\end{figure}

To augment this simple analysis, we also calculate the correlation between the article editing specialization of each language pair. 
We also employ Pearson correlation of log-transformed RCA profiles to quantify congruence in overall knowledge-production modes, in order to consider the interconnection between the editing portfolio and the knowledge production mode. The results also reveal coalitions among language editions depending on the genre (Figure~\ref{fig:heatmap_rca_combined_conspiracy}). The conspiracy genre demonstrates larger coalitions among less-prevalent languages in terms of viewership (bottom right). On the other hand, the science genre shows a large group of languages with high correlations in their specialization of knowledge production.  

\begin{figure}[htbp]
    \centering
    \includegraphics[width=0.8\linewidth]{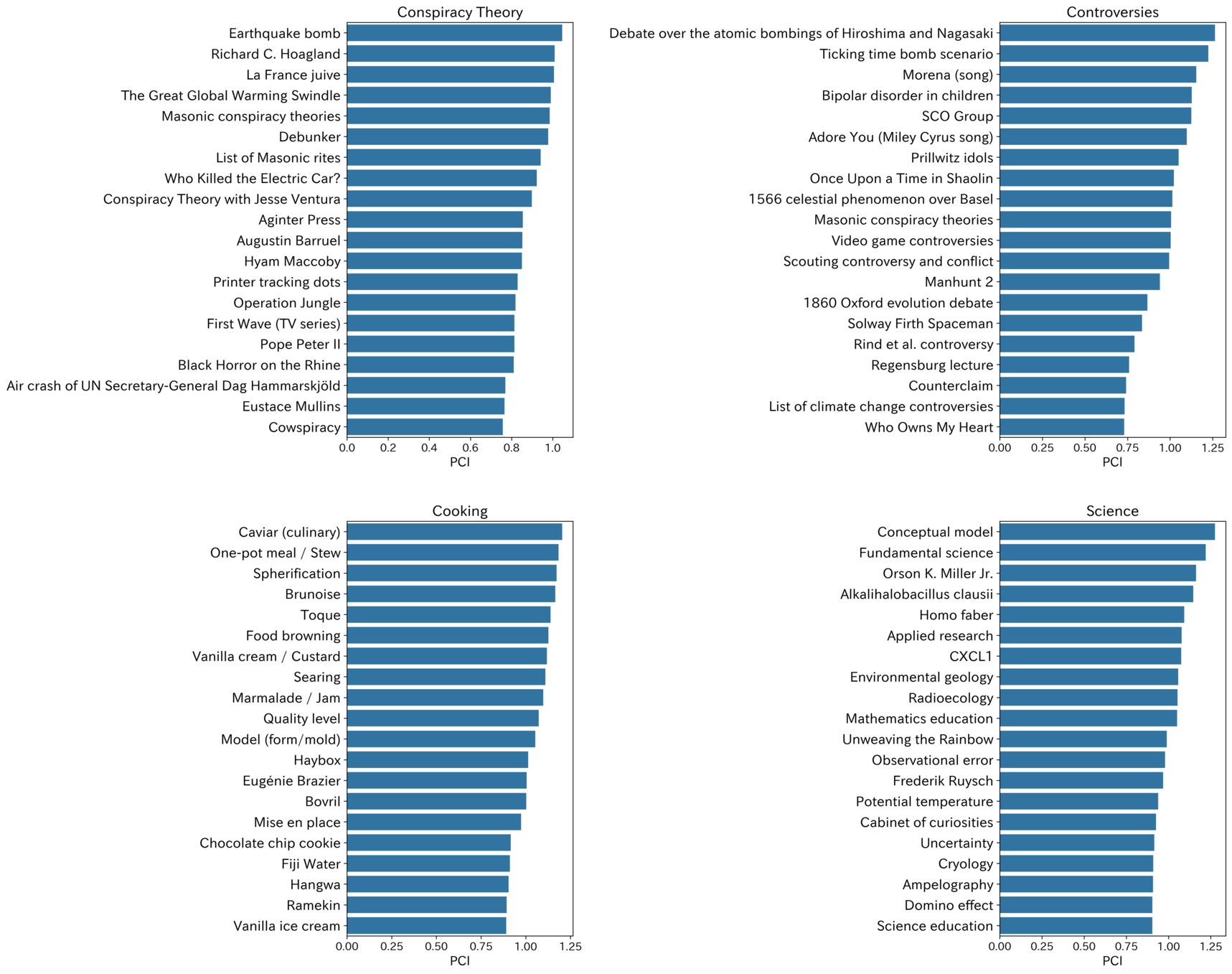}
    \caption{Top 20 Articles by Product Complexity Index (PCI). The charts display the top 20 articles with the highest PCI for each of the four genres. A high PCI signifies that an article requires sophisticated and rare production capabilities, meaning it is produced by a few, highly diversified language editions.}
    \label{fig:pci_top20_charts_conspi}
\end{figure}

Our language-agnostic analysis also detects articles specialized by specific languages, using the Product Complexity Index (PCI) (Method Sec.~\ref{sec:method_complexity}).
We rank articles by their PCI (Figure~\ref{fig:pci_top20_charts_conspi}). High-PCI articles demand sophisticated, specialized knowledge, reflecting their rarity and the complex capabilities of knowledge production.
Although there is some overlap in articles between the Conspiracy and Controversy genres, they feature different types of articles in the ranking.
The articles in the ranking of the Conspiracy genre are about a worldview that challenges official narratives, ranging from historical conspiracies to skepticism of modern science and government.
On the other hand, the list for the Controversy genre covers ``public conflict'' across all fields—history, science, and ethics—that relate to how modern society deals with controversial topics.
The lists for the Cooking and Science genres present not only articles on the main topics but also those on philosophical concepts or figures related to each genre. 

\begin{figure}[htbp]
    \centering
    \includegraphics[width=1\linewidth]{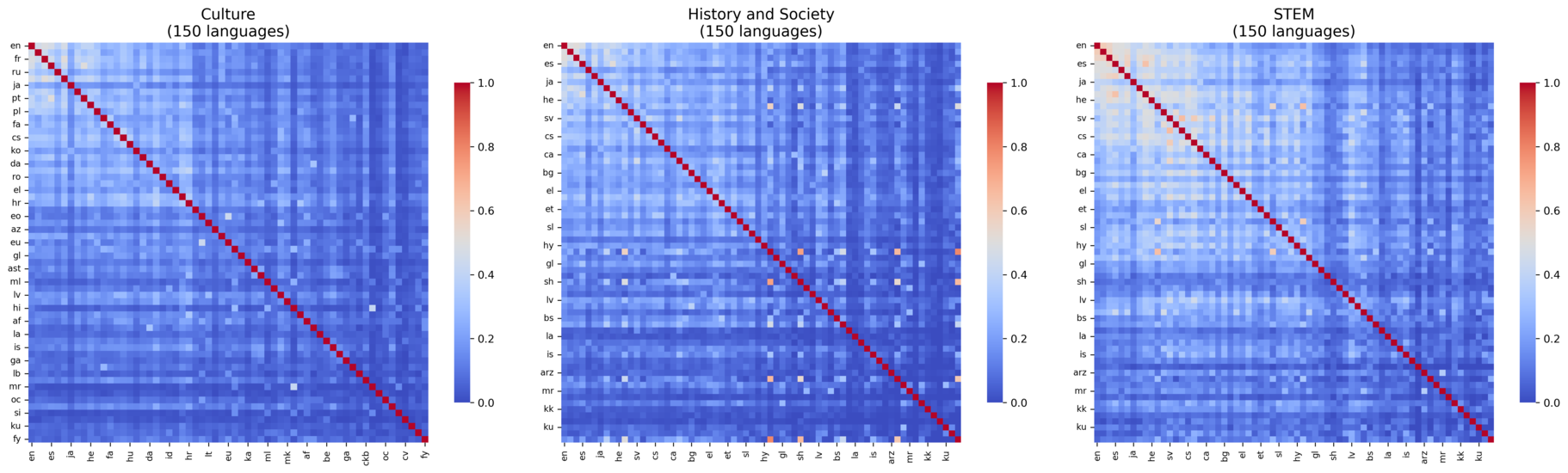}
    \caption{Heatmap of Pairwise Editorial Portfolio Similarity of each topic. The color represents the cosine similarity of editorial specialization between two languages; warmer colors indicate a higher degree of co-specialization on the same set of articles.}
    \label{fig:heatmap_combined_topic_parent_page}
\end{figure}

\begin{figure}[htbp]
    \centering
    \includegraphics[width=1\linewidth]{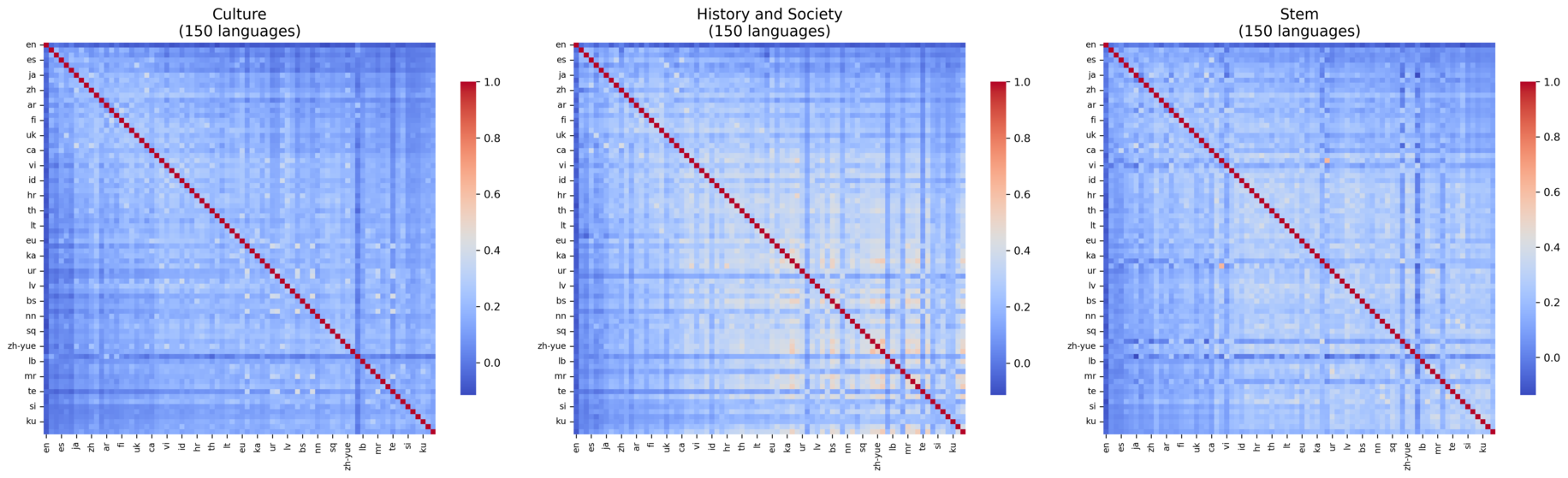}
    \caption{Heatmap of the Language RCA Similarity for Each Topic. Each heatmap visualizes the country's RCA similarity of knowledge production, calculated by the correlation between the RCA values of each country pair. Warmer colors represent higher values, suggesting that two languages have similar specializations in their knowledge production.}
    \label{fig:heatmap_rca_combined_topic_parent_page}
\end{figure}

These patterns may stem from the construction of genres, where we collect articles within specific categories of the Wikipedia platforms. For comparison, we leveraged the topics of Wikipedia articles that \cite{Valentim2021TrackingKP} identified, which contains 64 hierarchical topics (see Method Sec.~\ref{sec:method_topics}). 
We select meta-topics such as Culture, STEM, as well as History and Society. We refer to them as parent topics.
These three parent topics contain their subordinates as child topics. 
The analysis with the parent topics first finds that the STEM topics have higher similarity among the three, confirming our initial hypothesis (Figure~\ref{fig:heatmap_combined_topic_parent_page}).
We also study the analysis of Pearson correlation of log-transformed RCA profiles of the parent topic (Figure~\ref{fig:heatmap_rca_combined_topic_parent_page}). While not demonstrating distinctive patterns, we find a relatively high correlation among History and Society.

Since each parent topic covers millions of articles, their results become more evident when studying the results of 28 child topics (Figures~\ref{fig:heatmap_combined_topic_child_page_1}).
We find similar patterns to the Science genre (Figure~\ref{fig:sim_portfolio_genre_heatmap}) in Physics and Chemistry (Figure~\ref{fig:heatmap_combined_topic_child_page_1}).
We obtain qualitatively similar results in the correlation of RCA presented in (Figure~\ref{fig:heatmap_rca_combined_topic_child_page_1}) with some discrepancy with the finding in Figures~\ref{fig:heatmap_combined_topic_child_page_1}.
The RCA correlation captures a structure of shared specialization that is invisible to methods based on raw edit volume.
For example, in the "History" topic, the emergence of a highly correlated cluster in the bottom-right of Figure~\ref{fig:heatmap_rca_combined_topic_child_page_1} demonstrates that these language communities share a common mode of knowledge production, a relationship that the simpler cosine similarity of their edit portfolios does not capture (Figure~\ref{fig:heatmap_combined_topic_child_page_1}).
However, the results for Internet Culture, where editors can share common knowledge through the internet, also support our hypothesis that standardized topics or genres have higher similarity.

\begin{figure}[htbp]
    \centering
    \includegraphics[width=0.9\linewidth]{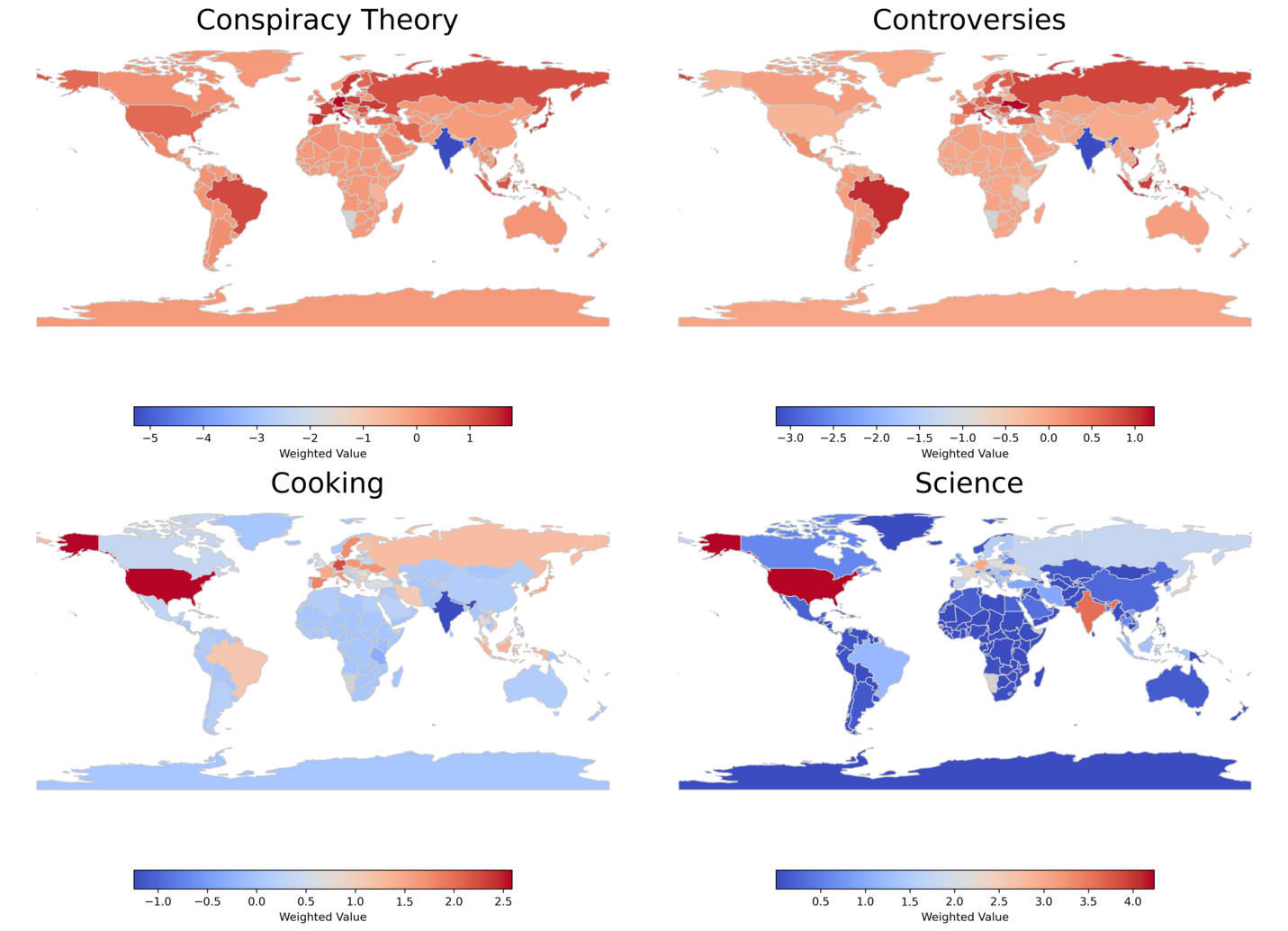}
    \caption{Geographic Distribution of the Economic Complexity Index (ECI) by Genre. The maps show the country-level ECI for each of the four genres. This metric is computed by weighting the ECI of each language edition by its viewership from each country. The maps reveal geographic variations in knowledge production complexity.}
    \label{fig:conspi_eci_index_map_by_category}
\end{figure}

\begin{figure}[htbp]
    \centering
    \includegraphics[width=1\linewidth]{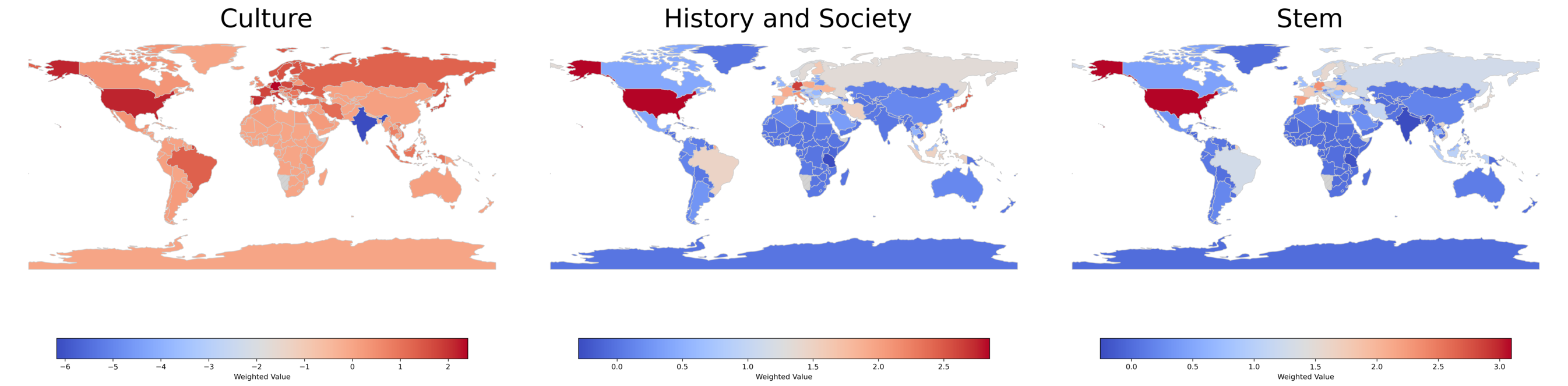}
    \caption{Geographic Distribution of the Economic Complexity Index (ECI) for Parent Topics. This figure maps the viewership-weighted ECI for the three broad parent topics: Culture, History and Society, and STEM. The Culture topic exhibits high complexity in regions with distinct linguistic spheres, reflecting the role of cultural context in knowledge production.}
    \label{fig:parent_eci_index_map_by_category}
\end{figure}

We next study the geopolitical distribution of knowledge production and are interested in the consumption of knowledge produced in each language version. 
Since access to Wikipedia language editions varies by country, we cannot directly translate the calculated complexity of a language into a country-level analysis.
In an extreme case, even if a language edition's articles have high complexity, it is insignificant if they have few accesses from a small number of countries.
To understand this, we weight the ECI of each language version by the total number of views from each country (Method Sec.~\ref{sec:method_view}). 
This allows us to calculate, for each country, the average complexity of articles that are "consumed" by its viewers.
We plot the analysis of the four genres on the map in Figure~\ref{fig:conspi_eci_index_map_by_category}, which demonstrates that the {Science} genre shows the lowest complexity, presumably because cultural or linguistic factors less skew interest in Science, as we discussed above. On the other hand, we find regional clusters with high complexity, such as around Europe in {Conspiracy Theory}. This means that those language editions exhibit distinct editing behavior in articles of the {Conspiracy Theory} genre. Notably, regions such as Europe and Asia (especially Japan) exhibit high complexity scores, consistent with patterns in economic complexity data~\cite{hidalgo2009building, hidalgo2021economic}. 
The analysis with the topics elucidates this finding: Culture topics exhibit high complexity where linguistic factors play a pivotal role (Figure~\ref{fig:parent_eci_index_map_by_category}). 
We also decomposed the parent topic-based analysis into a child topic-based analysis (Figure~\ref{fig:map_children_combined_grid_combined}). Tables~\ref{tbl:country_broader} and \ref{tbl:country_selected} show that the complexity ranking can vary among topics or genres. More specifically, we find that the ranking of genre-based analysis is subject to change across the genres (Figure~\ref{fig:bumpy_ranking_conspi}), but we have a relatively stable ranking across the topics  (Figure~\ref{fig:bumpy_eci_ranking_parent}). 
 
The consumption of this complexity varies among topics, suggesting domains play a pivotal role. For example, we find high similarity between Physics and Chemistry, which suggests standardized knowledge production. However, we also find those topics have high complexity on the map (Figure~\ref{fig:map_children_combined_grid_combined}). The viewers in each country consume articles that have high complexity (i.e., are highly specific to that language). Interestingly, the map demonstrates the relatively low complexity of the Mathematics topic in global consumption. Given that the similarity among languages is high (Figure~\ref{fig:heatmap_combined_topic_child_page_1} and ~\ref{fig:heatmap_rca_combined_topic_child_page_1}), this result implies that we have a similar capability of mathematical knowledge production and our consumption of that knowledge is also uniform~\footnote{We consider this result is mainly because articles on Mathematics topics often contain articles about numbers, such as the article on ``2000 (number).''}. 
This indicates that even within similar genres or topics, the relationships between production and consumption can differ.

Our findings reveal that the mode of knowledge production on Wikipedia reflects complex interactions among languages and their domains. 
The analysis identifies segmented language groups that mirror cultural, geopolitical, and economic contexts. Notably, languages with large user bases often specialize in articles differently than smaller or regionally focused languages. These results suggest that collaborative norms and broader factors related to language and geographical location, including geopolitical considerations, shape online knowledge ecosystems and editorial preferences. This complex structure may explain why the platform does not exhibit specific trends over the years in their knowledge production capabilities. For example, our calculation of economic complexity among languages using annual data shows stable similarity values (Figure~\ref{fig:yearly_heatmap_trend_STEM}). While having fluctuations within Conspiracy genres (Figure~\ref{fig:yearly_heatmap_trend_conspiracy_theory}), these do not indicate specific trends.
Additionally, we also find that the structure of knowledge production does not predict new article creations well.
We employed relatedness density to predict article creations following the original complexity analysis~\cite{hidalgo2009building, hidalgo2021economic}, but we found a downward trend in AUC despite the training data expanding in later years (Figure~\ref{fig:related_ness_prediction_conspi} and \ref{fig:related_ness_prediction_parent}).
This downward trend is common in child topics (Figure~\ref{fig:related_ness_prediction_child}), implying that this is a platform-level phenomenon. This suggests that the structure of knowledge production becomes more complex as time passes, and user activity cannot be explained solely by activities within the platform.

\begin{figure}[htbp]
    \centering
    \includegraphics[width=\linewidth]{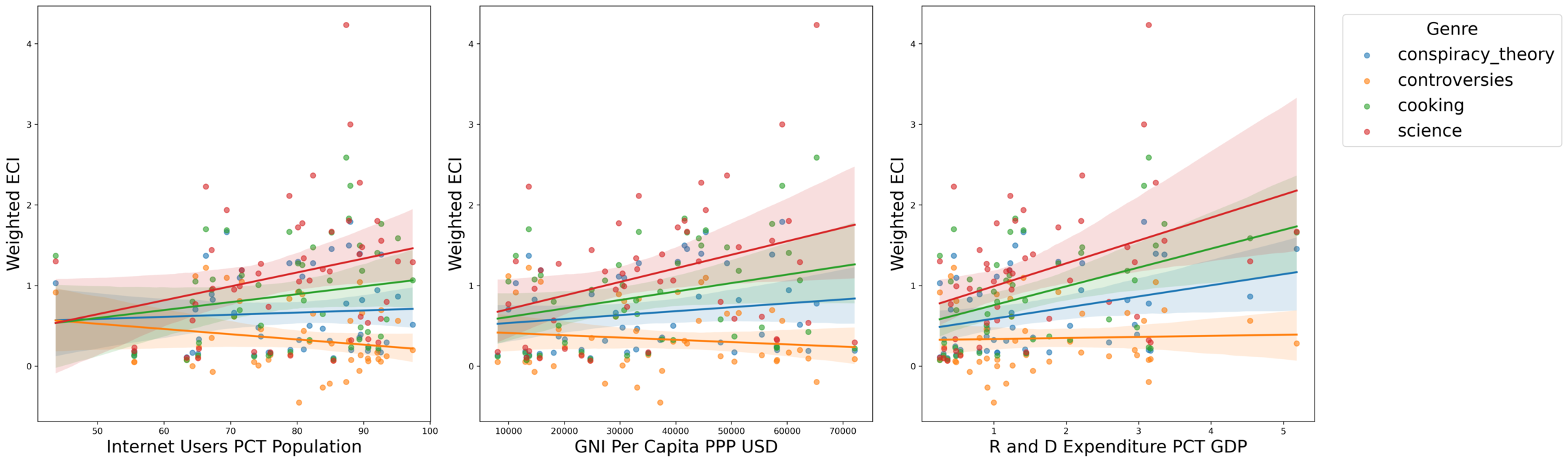}
    \caption{ The relationship is shown between a country's viewership-weighted ECI in the Conspiracy Theory genre and various economic indicators, including Internet users (percent of the population), GNI per capita, and R\&D expenditure (percent of GDP). Each point represents a country, and the figures focus on the top 50 countries by weighted ECI.
    }
    \label{fig:economic_index_regression_conspiracy_theory}
\end{figure}

\begin{figure}[htbp]
    \centering
    \includegraphics[width=\linewidth]{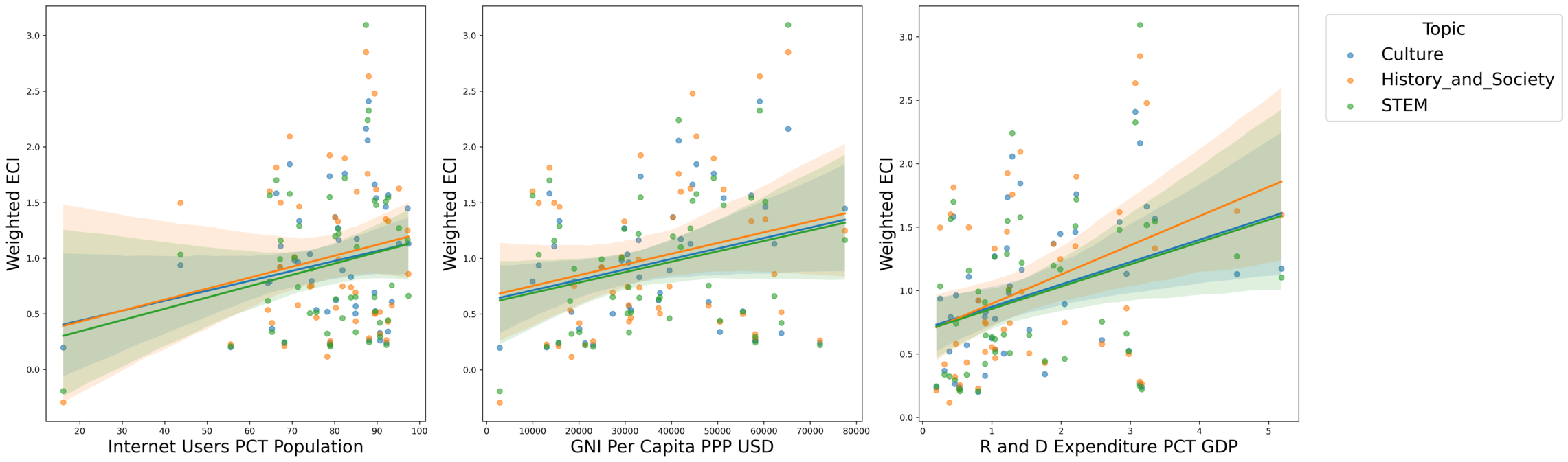}
    \caption{ 
    The relationship is shown between a country's viewership-weighted ECI in the Culture topic and various economic indicators, including Internet users (percent of the population), GNI per capita, and R\&D expenditure (percent of GDP). Each point represents a country, and the figures focus on the top 50 countries by weighted ECI.
    }
    \label{fig:economic_index_regression_culture}
\end{figure}

We then turn our interest to investigating how such structural differences are associated with external factors that can affect editorial or viewership behavior. For this, we regress the ECI on economic indices, inspired by the original economic complexity analysis~\cite{hidalgo2009building, hidalgo2021economic}.
We collected the economic index data from~\cite{WorldBankOpenData} and found that some indices, such as GNP per capita or Research and Development (R\&D), correlate with the ECI (Figures~\ref{fig:economic_index_regression_conspiracy_theory} and \ref{fig:economic_index_regression_culture}). The regression results from the genre-based analysis reveal that the Science genre correlates well with the R\&D index, indicating that the editorial activity of some genres can outperform others in their associations with economic activities.

\section{Discussion and Conclusion}

This study proposed a new lens through which to view the structure of knowledge production across languages and the globe, leveraging economic complexity analysis on the Wikipedia platform. Our findings revealed that the structure is not flat; different language communities develop distinct capabilities for producing different genres or topics of knowledge. We demonstrated a sharp contrast between the standardized, uniform production of scientific knowledge and the fragmented, culturally-bound production of topics like history and conspiracy theories. The detected structural differences mirror geopolitical and linguistic boundaries, suggesting that our collective knowledge ecosystem is deeply shaped by real-world social and political contexts.

The implications of this structure of ``knowledge production'' are far-reaching. Our map of knowledge capabilities reveals which communities are central to producing certain types of information and which are on the periphery. This highlights potential vulnerabilities, such as the risk of a few dominant language communities defining what constitutes ``global'' knowledge. 
AI models like LLMs, which learn from Wikipedia, can inherit the structural imbalances we detected in this study. This means that while these AI systems may have standardized knowledge in topics like Science or STEM, they might also have geopolitical biases in more culturally-specific topics.

However, this study has limitations, as do other studies. We base our analysis solely on Wikipedia, so our findings might not apply to other knowledge platforms. We use the number of edits as a proxy for knowledge production, which is a simplification because not all edits have equal value. The way we group articles into genres could also influence the results. Finally, the correlations we find between knowledge complexity and economic factors do not necessarily mean one causes the other. Future research should address these points to build a more complete picture.

\section{Data \& Method}
This section discusses the construction of the dataset and the method for this study.

\subsection{Wikipedia Data}\label{sec:method_topics}
We first collect Wikipedia editing history data from 2001 to 2024 for over 150 language editions, excluding bot contributions. Building on the publicly available MediaWiki API~\footnote{\url{https://www.mediawiki.org/wiki/API:Main_page}} we identify articles in each language edition that appear in at least one high-level category (\texttt{Conspiracy Theories}, \texttt{Wikipedia Controversial Topics}, \texttt{Cooking}, and \texttt{Natural Science}). 
We then perform a cross-lingual search by tracing interlanguage links to ensure the inclusion of articles possibly categorized differently in another language. It should be noted that category assignments can vary across languages, and some articles identified in the procedure may not belong to the same category in that version, as humans assign categories. 
For example, even if an article titled ``Hamburger'' is found under the Cooking category in English Wikipedia, this does not guarantee that ``Hamburger'' in another language belongs to the Cooking category in that version, as humans assign categories.
To prevent misunderstandings regarding the term ``category,'' which is official Wikipedia terminology, we label these four sets of articles obtained through this procedure as \textbf{Genre}. We construct four sets of articles, referring simply to them as \texttt{Conspiracy Theory}, \texttt{Controversy}, \texttt{Cooking}, and \texttt{Science}. For each set of articles, we retrieve the complete revision histories of editors who have contributed to these genres at least once, capturing their broader editing patterns, including edits beyond the targeted genres. The full set of editing histories of the users results in over 36.5 million users and around 450,000 titles. In addition to this genre dataset, we construct a dataset that can represent the Wikipedia platforms in general. We leverage the dataset constructed by \cite{Valentim2021TrackingKP} that tracks article creation dynamics across different language editions with the estimated topic labels. We conduct the same analysis with those data to capture the general view of the platforms. 

\subsection{Viewer data by Countries}~\label{sec:method_view}
To compute the viewer weights for each country–language pair, we first collect access data for every language edition from every country via the MediaWiki API~\footnote{\url{https://wikimedia.org/api/rest\_v1/metrics/pageviews/top-by-country}}).
We normalize the weights so that, within each country, they sum to 1.
This data is only available for 2016 statistics, and we use the data from 2016 for the analysis with this constructed weight.

\subsection{Economic Complexity Analysis}~\label{sec:method_complexity}
To quantify the knowledge production capabilities of different language communities, we adapt the framework of economic complexity, originally developed to analyze the industrial capabilities of countries based on their export data~\cite{hidalgo2009building, hidalgo2021economic}. In our context, we create an analogy where languages are equivalent to countries, and Wikipedia articles are equivalent to products. The number of edits an article receives in a specific language serves as a proxy for its ``production'' volume.
Note that we conduct country-level analysis, but it is based on the weighted value of this language-level analysis using country-level Wikipedia access data.

We determine whether a language $l$ has a specialization in producing an article $a$. This is quantified using the concept of Revealed Comparative Advantage (RCA). We construct a binary matrix $M_{la}$, where $M_{la} = 1$ if language $l$ has a comparative advantage in article $a$, and $M_{la} = 0$ otherwise. We consider Language $l$ to have a comparative advantage if its share of edits on an article is greater than the share of edits on that same article across all languages. Formally, we set $M_{la} = 1$ if $RCA_{la} \ge 1$, where we define $RCA_{la}$ as:
$$
RCA_{la} = \frac{E_{la} / \sum_{a'} E_{la'}}{\sum_{l'} E_{l'a} / \sum_{l', a'} E_{l'a'}}
$$
Here, $E_{la}$ represents the number of edits on article $a$ in language $l$. This matrix $M_{la}$ forms the foundation for our complexity calculations, indicating which languages specialize in which articles.

The presented paper first calculates the cosine similarity of vectors that contain the number of revisions in each article to study the similarity among language editions.
Then, we also calculate the pairwise similarities in knowledge production in terms of RCA. The similarity calculations involve cosine similarity among the portfolios of RCA for articles, which captures the languages' specialization in knowledge production.
Following~\cite{bahar2014neighbors}, we calculate the Pearson correlation between the log-transformed RCA vectors of two languages for the similarity measurement. A higher value of the index thus signifies that the two countries have more similar language production modes in the sense that they have similar articles with similar levels of comparative advantage. We derive the Economic Complexity Index (ECI) by iteratively capturing the interplay between a language’s breadth of specialized articles and the rarity of those articles across all languages. The standardized language complexities (mean zero, unit variance) constitute the ECI, whereby a high ECI signals a community capable of producing a wide array of exclusive, sophisticated content.

\renewcommand{\thesection}{S\arabic{section}}
\renewcommand{\thefigure}{S\arabic{figure}}
\renewcommand{\thetable}{S\arabic{table}}
\setcounter{figure}{0}
\setcounter{table}{0}

\section*{Supplemental Information}
\begin{figure}[htbp]
    \centering
    \includegraphics[width=0.7\linewidth]{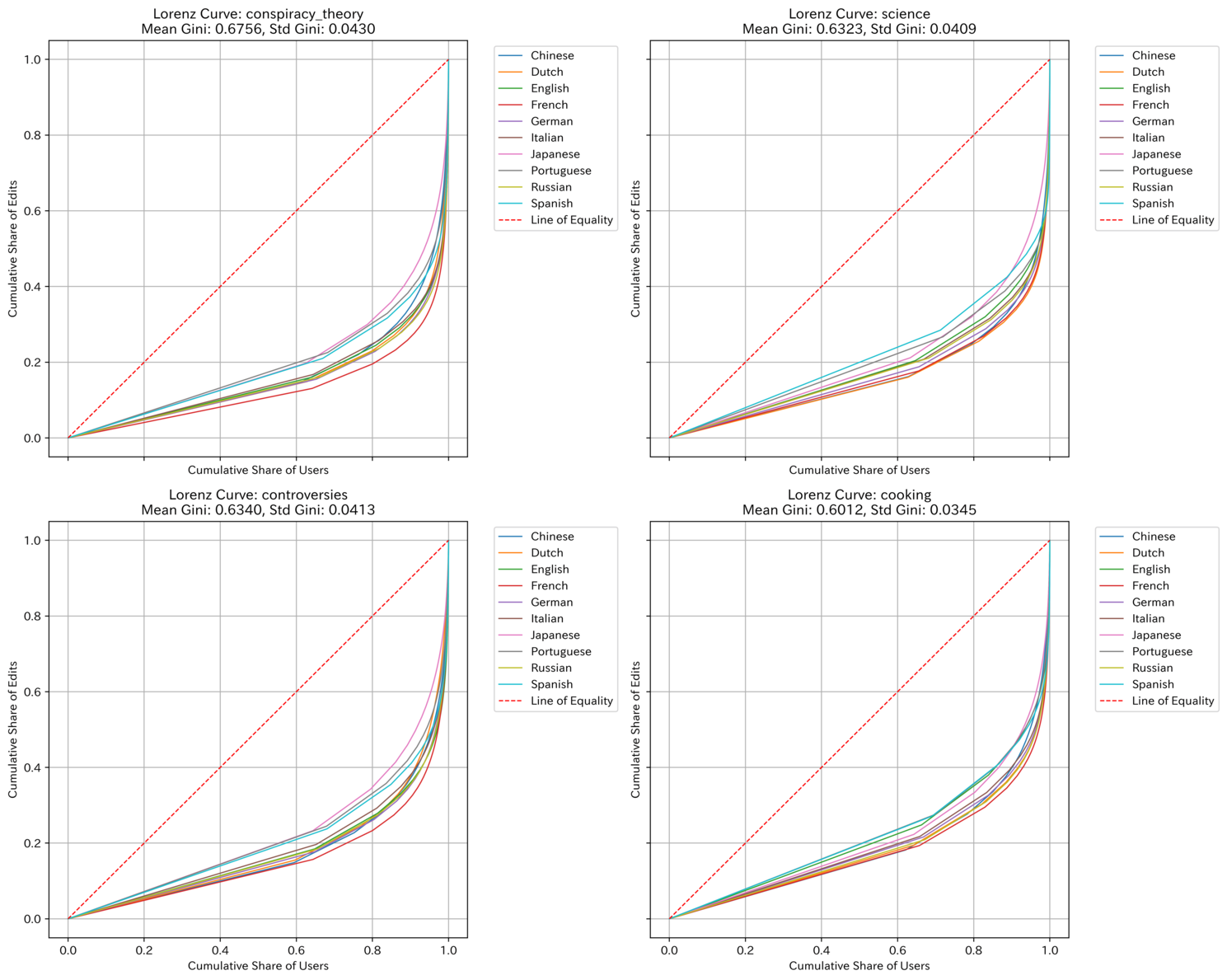}
    \caption{Lorenz curves showing the concentration of editorial activity for the top 10 languages. The x-axis represents the cumulative percentage of editors, and the y-axis represents the cumulative percentage of edits. The diagonal line signifies perfect equality.}
    \label{fig:lorenz_curve}
\end{figure}

\begin{figure}[htbp]
    \centering
    \includegraphics[width=0.7\linewidth]{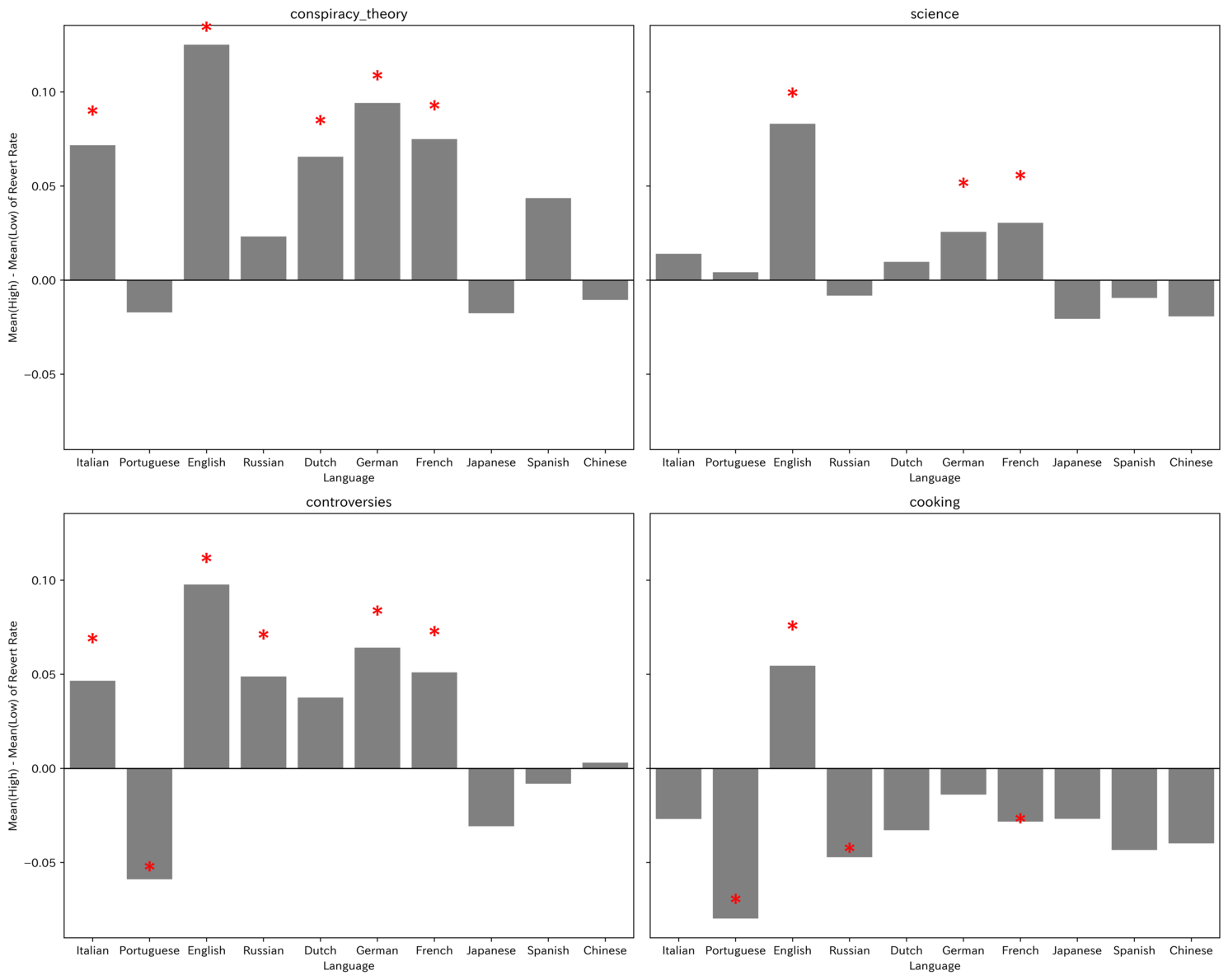}
    \caption{Difference in edit reversion rates between high- and low-engagement editors. We define engagement by the number of edits per article (top 50\% vs. bottom 50\%). Positive values indicate that high-engagement editors have a higher revert rate. $^*$ represents $\text{p-val} < 0.01$.}
    \label{fig:revert_rate_comp_bar}
\end{figure}

\begin{figure}[htbp]
    \centering
    \includegraphics[width=0.5\linewidth]{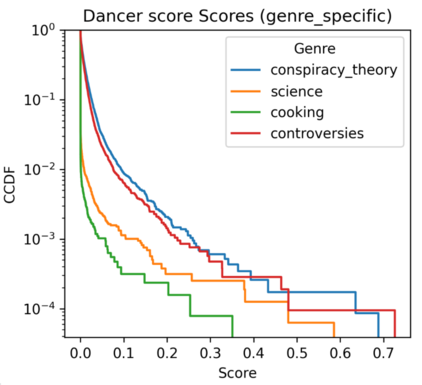}
    \caption{Semantic diversity in consecutive edits, which we measure with the Dancer Score inspired by~\cite{DaleArchitectural2024}. A higher score indicates that an editor tends to work on more semantically diverse articles in a single session. Editors in Conspiracy Theory and Controversy exhibit broader editing patterns, while those in Cooking and Science show more specialized activity, focusing on semantically related articles.}
    \label{fig:dancer_score_rate}
\end{figure}

\begin{figure}[htbp]
    \centering
    \includegraphics[width=1\linewidth]{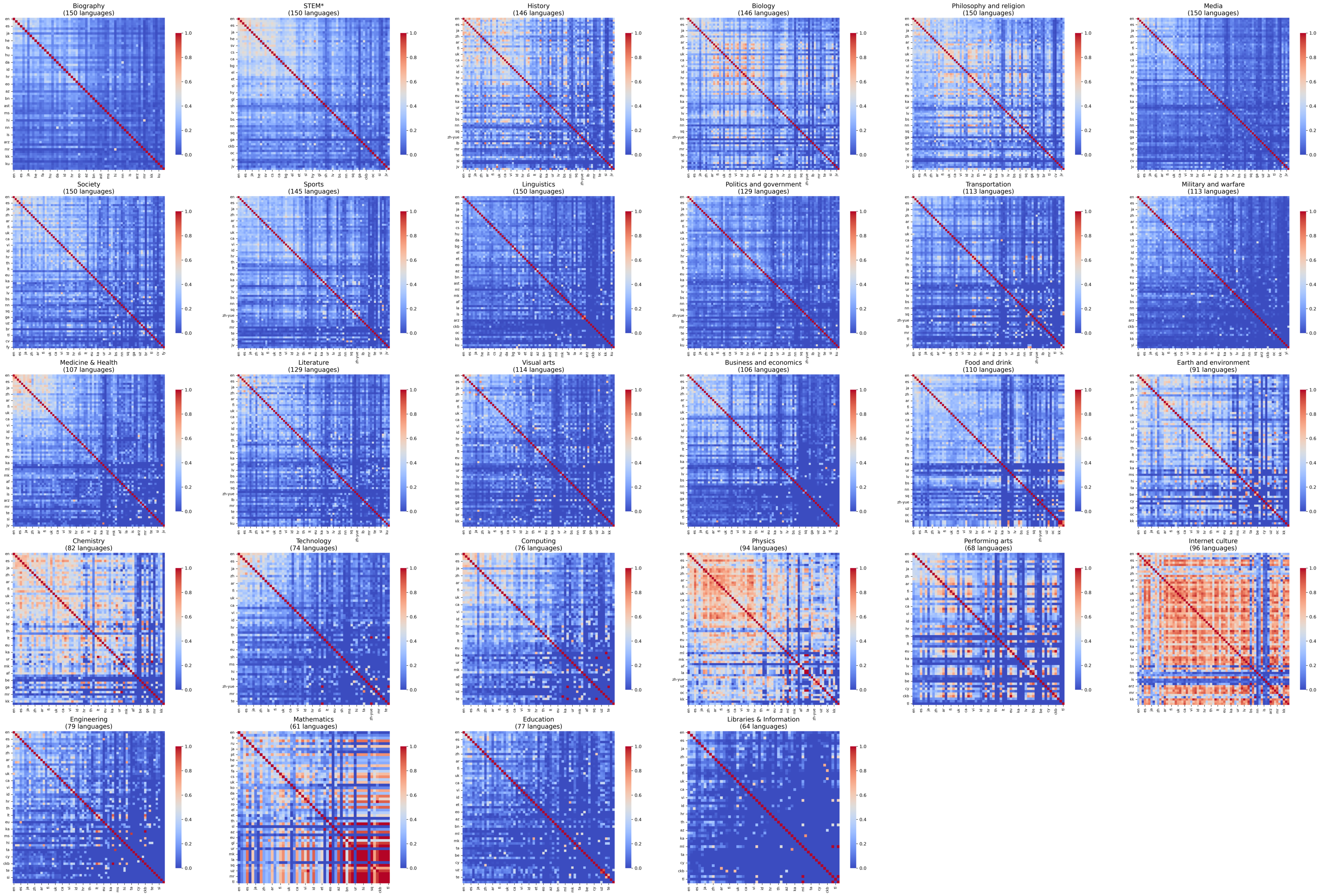}
    \caption{
    Heatmap of Pairwise Editorial Portfolio Similarity for 28 individual child topics. Warmer colors represent the cosine similarity of editorial specialization between two languages, indicating a higher degree of co-specialization on the same set of articles.}
    \label{fig:heatmap_combined_topic_child_page_1}
\end{figure}

\begin{figure}[htbp]
    \centering
    \includegraphics[width=1\linewidth]{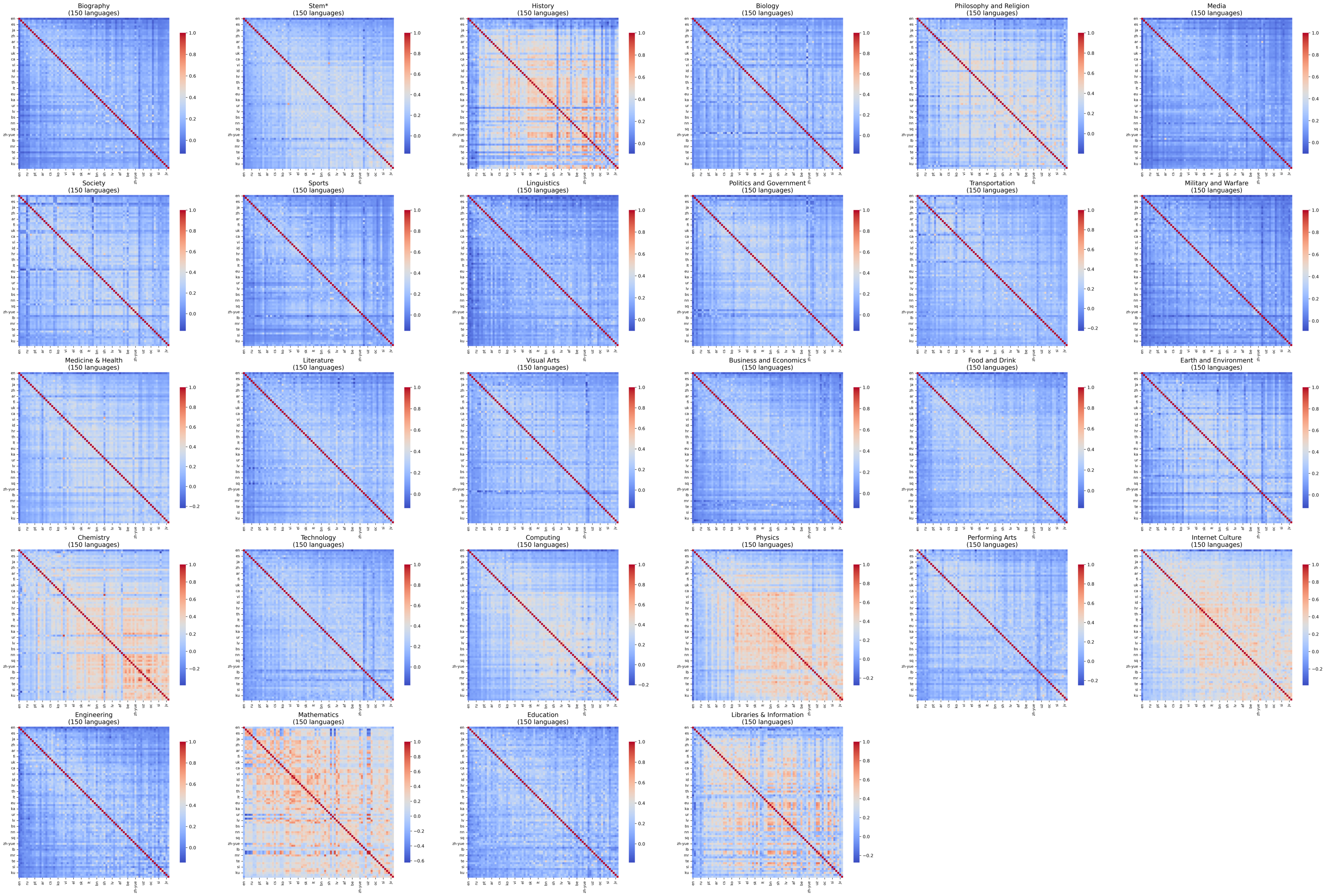}
    \caption{
    Heatmap of the Language RCA Similarity for 28 individual child topics. Each heatmap visualizes the countries' RCA similarity of knowledge production, calculated from the correlation between the RCA values of each country pair. Warmer colors represent higher values, suggesting that two languages have similar specialization in their knowledge production.}
    \label{fig:heatmap_rca_combined_topic_child_page_1}
\end{figure}

\begin{figure}[htbp]
    \centering
    \includegraphics[width=1\linewidth]{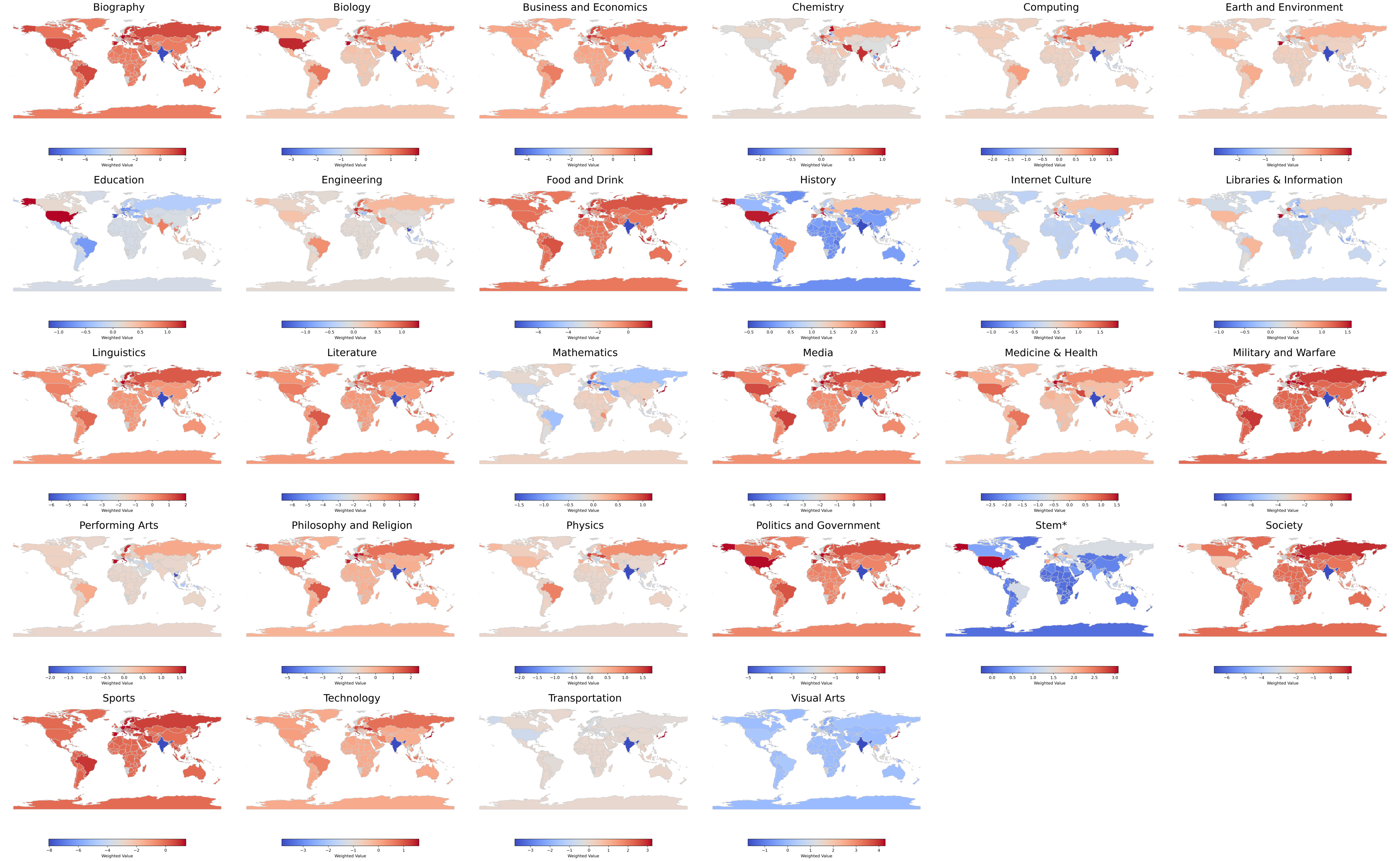}
    \caption{Geographic distribution of the Economic Complexity Index (ECI) for child topics. We weight each country's ECI by the pageviews from that country to each language edition. The geographic distribution of complexity varies significantly by topic.}
    \label{fig:map_children_combined_grid_combined}
\end{figure}

\begin{figure}[htbp]
  \centering
  \begin{subfigure}[b]{0.48\linewidth}
    \centering
    \includegraphics[width=\linewidth]{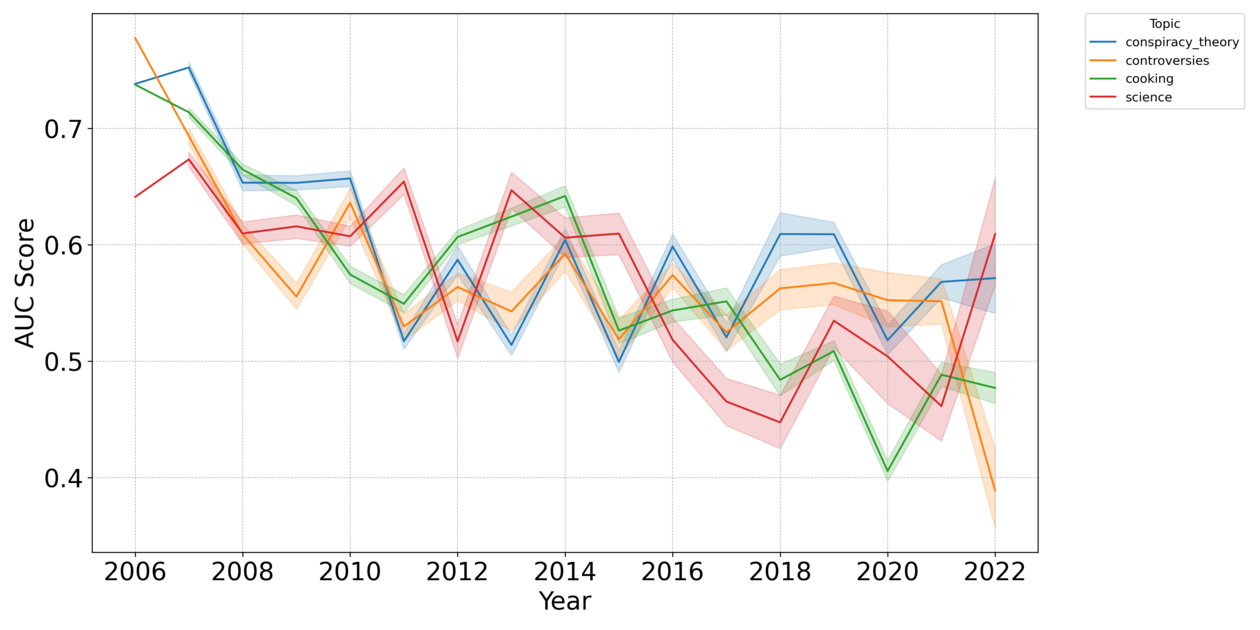}
    \caption{Conspiracy Genres}
    \label{fig:related_ness_prediction_conspi}
  \end{subfigure}
  \hfill
  \begin{subfigure}[b]{0.48\linewidth}
    \centering
    \includegraphics[width=\linewidth]{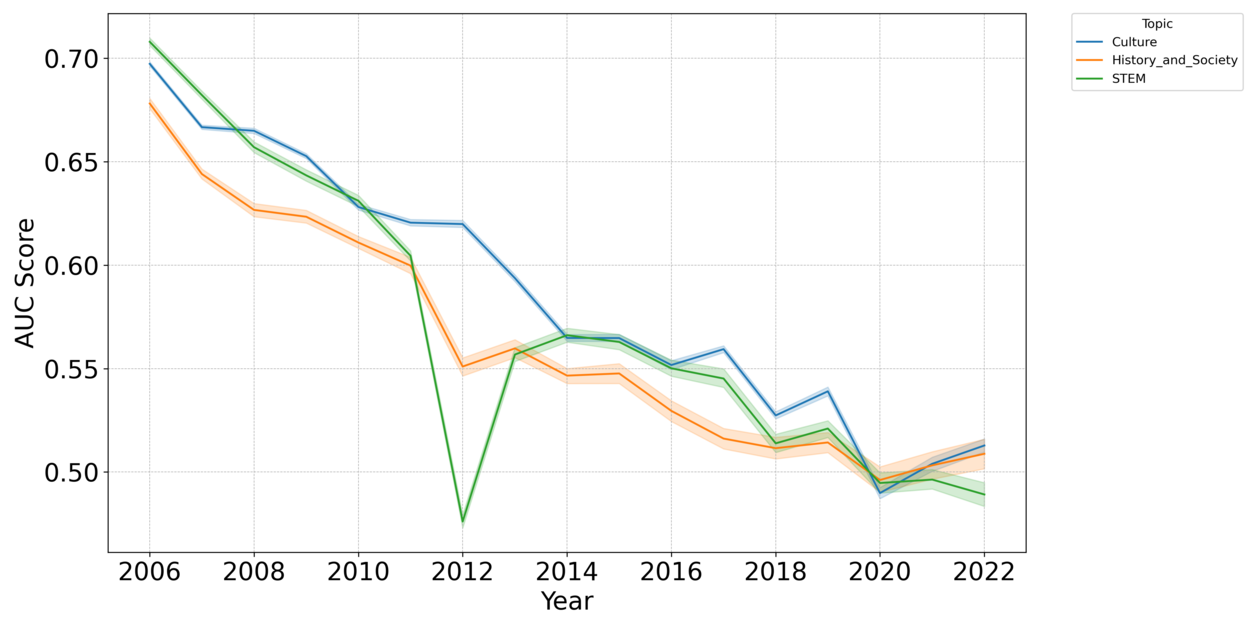}
    \caption{Parent Topics}
    \label{fig:related_ness_prediction_parent}
  \end{subfigure}
  \caption{The area under the curve (AUC) for predictions of new-article creation over time based on relatedness density, using logistic regression; (a) The genres, (b) the parent topics. We observe a consistent downward trend in AUC across all categories. This suggests that as Wikipedia matures, new knowledge creation becomes less predictable solely from the existing structure of knowledge production capabilities.}
  \label{fig:relatedness_comparison}
\end{figure}

\begin{figure}[htbp]
    \centering
    \includegraphics[width=\linewidth]{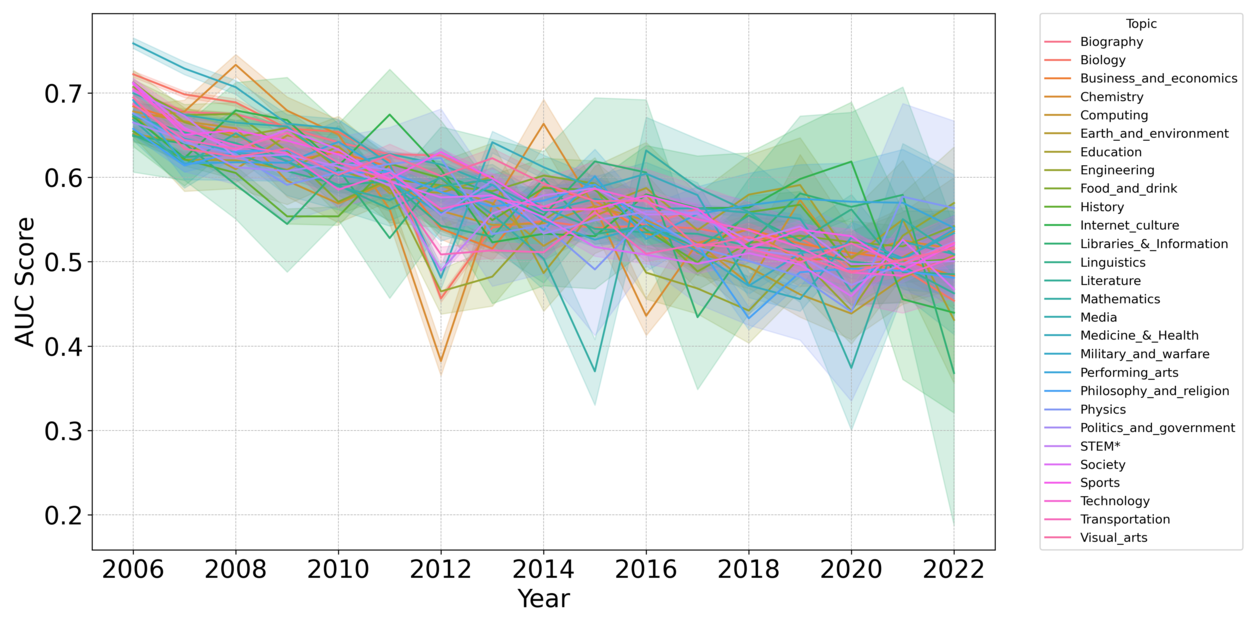}
    \caption{The area under the curve (AUC) for predictions of new-article creation over time based on relatedness density for child topics using the same settings as Figure~\ref{fig:relatedness_comparison}.}
    \label{fig:related_ness_prediction_child}
\end{figure}

\begin{figure}[htbp]
    \centering
    \includegraphics[width=1\linewidth]{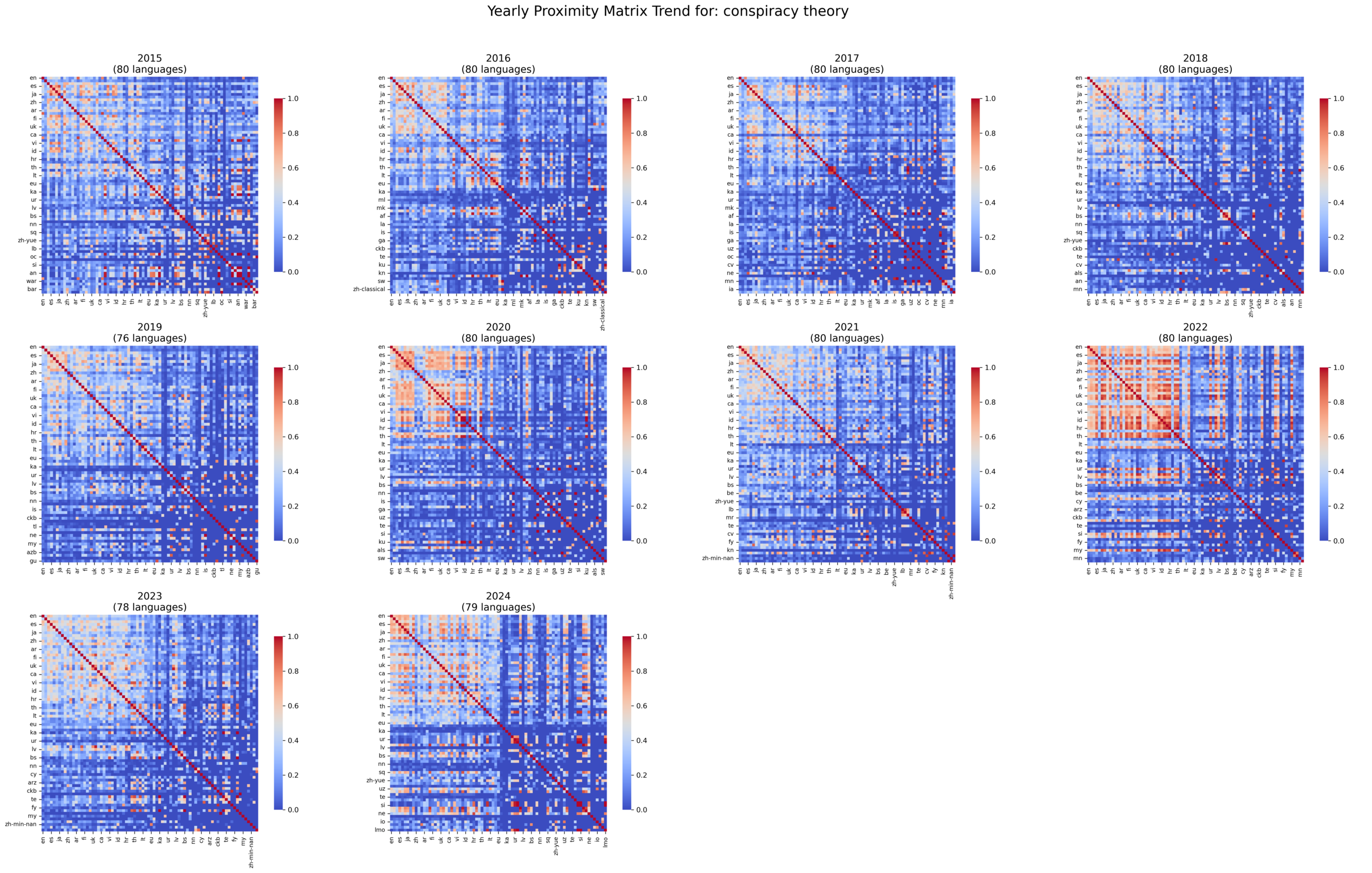}
    \caption{Temporal evolution of the proximity structure in the Conspiracy Theory genre. Each heatmap represents results calculated from one year of data.}
    \label{fig:yearly_heatmap_trend_conspiracy_theory}
\end{figure}

\begin{figure}[htbp]
    \centering
    \includegraphics[width=1\linewidth]{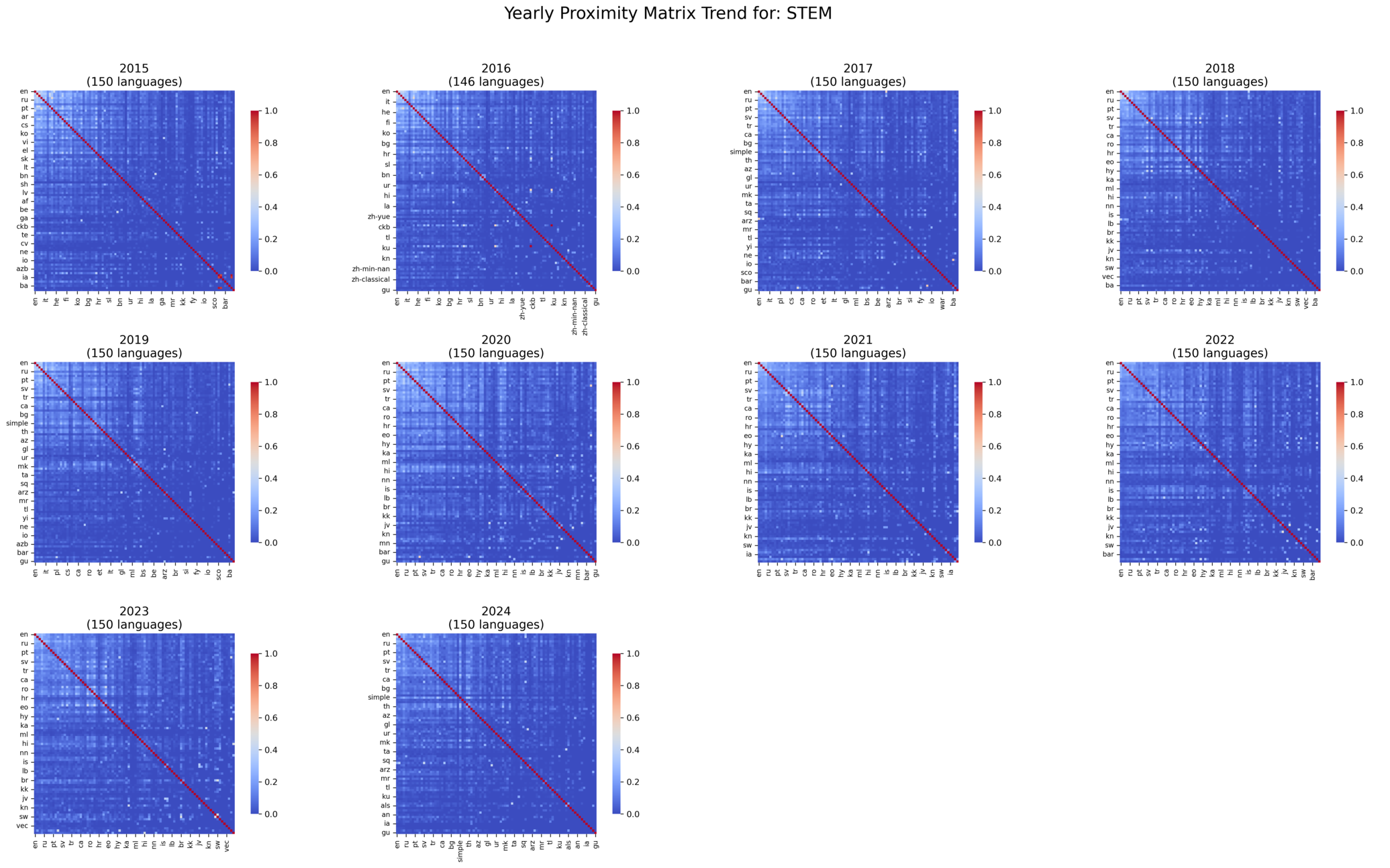}
    \caption{Temporal evolution of the proximity structure for a parent topic (STEM). The knowledge production capability for STEM is relatively stable over time, indicating that editors produce and maintain science-related knowledge on the platform in a globally standardized manner that has not significantly changed.}
    \label{fig:yearly_heatmap_trend_STEM}
\end{figure}

\begin{figure}[htbp]
    \centering
    \includegraphics[width=1\linewidth]{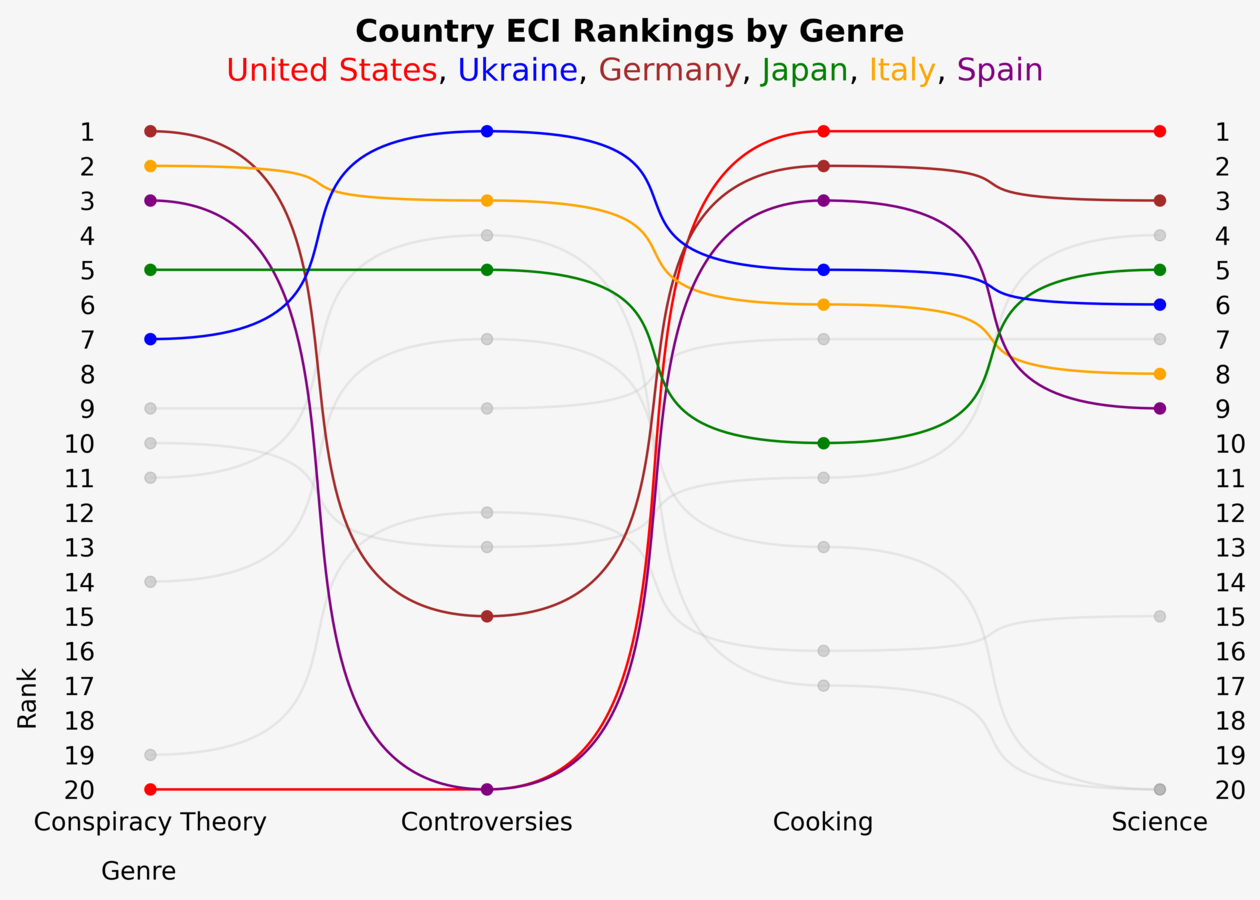}
    \caption{Ranking Transition among the top 20 countries by ECI for the Conspiracy genre. The complete ranking table appears in Table~\ref{tbl:country_selected}.}
    \label{fig:bumpy_ranking_conspi}
\end{figure}

\begin{figure}[htbp]
    \centering
    \includegraphics[width=1\linewidth]{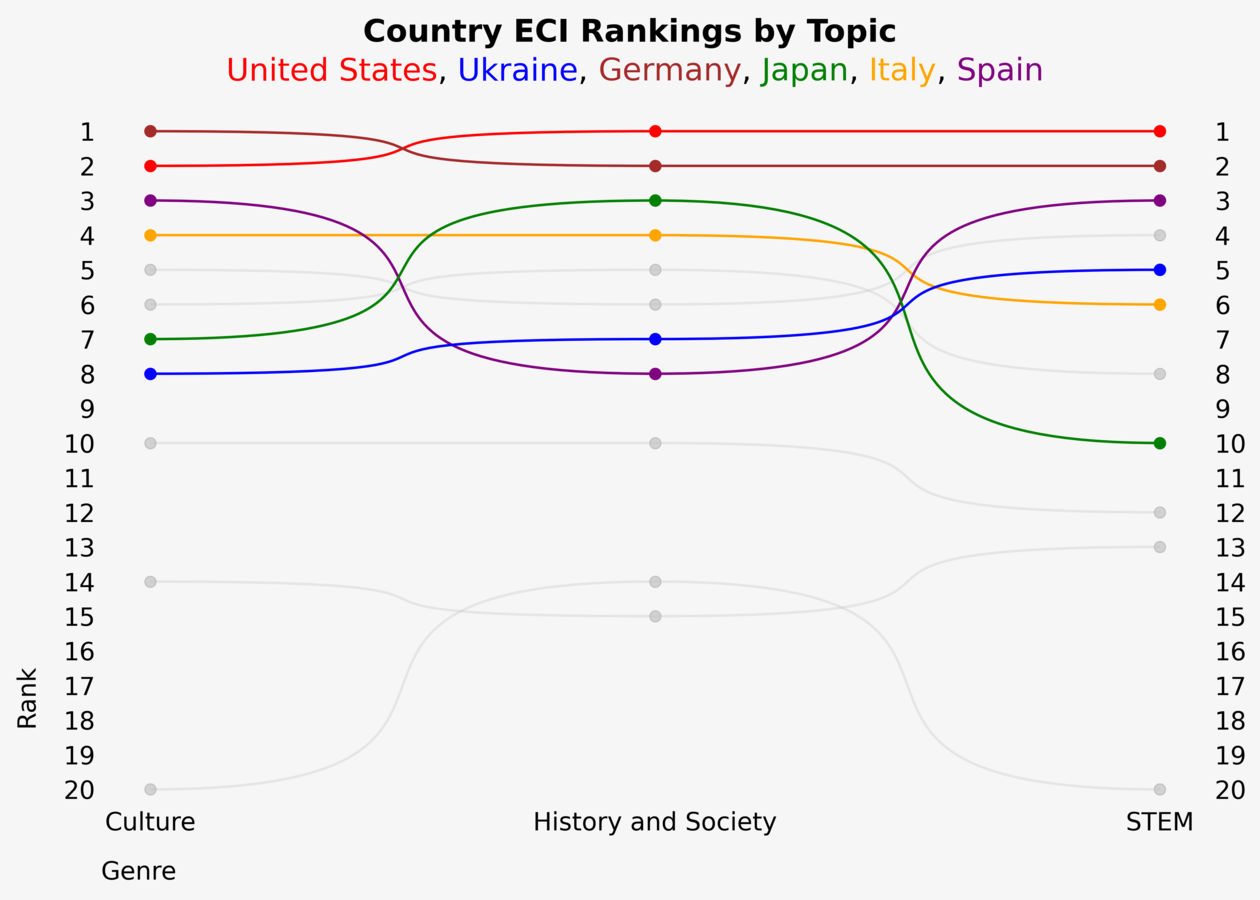}
    \caption{Ranking Transition among the top 20 countries by ECI for the Culture topic. We present the complete ranking table in Table~\ref{tbl:country_broader}.}
    \label{fig:bumpy_eci_ranking_parent}
\end{figure}

\clearpage

\begin{table}[htbp]
    \centering \small
    \caption{Top 20 country rankings by viewership-weighted Economic Complexity Index (ECI) in four selected genres. The leading countries vary significantly by genre, e.g., Germany in the Conspiracy Theory genre and the United States in the Science genre.}
    \label{tbl:country_selected}
    \begin{tabular}{lllll}
        \toprule
         & Conspiracy Theory & Controversies & Cooking & Science \\
        \midrule
        1 & Germany & Ukraine & United States & United States \\
        2 & Italy & Viet Nam & Germany & India \\
        3 & Spain & Italy & Spain & Germany \\
        4 & Israel & Brazil & Sweden & France \\
        5 & Japan & Japan & Ukraine & Japan \\
        6 & Sweden & Czechia & Italy & Ukraine \\
        7 & Ukraine & Indonesia & Poland & Poland \\
        8 & Czechia & Russian & Israel & Italy \\
        9 & Poland & Poland & Korea & Spain \\
        10 & France & Romania & Japan & Netherlands \\
        11 & Brazil & Sweden & France & Russian \\
        12 & Russian & Finland & Netherlands & Czechia \\
        13 & Romania & France & Indonesia & Israel \\
        14 & Indonesia & Türkiye & Czechia & Sweden \\
        15 & Netherlands & Germany & Russian & Finland \\
        16 & Bulgaria & Korea & Finland & Bulgaria \\
        17 & Korea & Serbia & Brazil & Estonia \\
        18 & Iran & Hungary & Iran & Hungary \\
        19 & Finland & Bulgaria & Denmark & Korea \\
        20 & United States & Spain & Malaysia & Indonesia \\
        \bottomrule
    \end{tabular}
\end{table}

\begin{table}[htbp]
    \centering \small
    \caption{Top 20 country rankings by viewership-weighted Economic Complexity Index (ECI) in three broad parent topics. The table shows which countries lead in the consumption of complex knowledge in each domain.}
    \label{tbl:country_broader}
    \begin{tabular}{llll}
        \toprule
         & Culture & History and Society & STEM \\
        \midrule
        1 & Germany & United States & United States \\
        2 & United States & Germany & Germany \\
        3 & Spain & Japan & Spain \\
        4 & Italy & Italy & France \\
        5 & France & Poland & Ukraine \\
        6 & Poland & France & Italy \\
        7 & Japan & Ukraine & Viet Nam \\
        8 & Ukraine & Spain & Poland \\
        9 & Sweden & Korea & Sweden \\
        10 & Finland & Finland & Japan \\
        11 & Netherlands & Viet Nam & Netherlands \\
        12 & Norway & Israel & Finland \\
        13 & Czechia & Iran & Brazil \\
        14 & Brazil & Indonesia & Korea \\
        15 & Russian & Brazil & Russian \\
        16 & Israel & Czechia & Hungary \\
        17 & Hungary & Netherlands & Czechia \\
        18 & Denmark & Sweden & Norway \\
        19 & Korea & Russian & Iran \\
        20 & Iran & Norway & Israel \\
        \bottomrule
    \end{tabular}
\end{table}

\end{document}